\documentclass[letterpaper,comsoc]{IEEEtran}

\usepackage[T1]{fontenc}
\usepackage{graphicx}
\usepackage{subfigure}
\usepackage{graphics}
\usepackage{algorithm}
\usepackage{algorithmic}
\usepackage{amssymb}
\usepackage{amsthm}

\usepackage[algo2e,linesnumbered,ruled]{algorithm2e}

\ifCLASSINFOpdf
\else
\fi
\usepackage{amsmath}

\usepackage[cmintegrals]{newtxmath}

\begin{document}

\title{On-Demand Density-Aware UAV Base Station 3D Placement for Arbitrarily Distributed Users with Guaranteed Data Rates}


\author{Chuan-Chi Lai,~\IEEEmembership{Member,~IEEE,} Chun-Ting Chen, and Li-Chun Wang*,~\IEEEmembership{Fellow,~IEEE}%
\IEEEcompsocitemizethanks{
	\IEEEcompsocthanksitem This research is supported by Ministry of Science and Technology, Taiwan, R.O.C. under the Grant MOST 106-2622-8-009-017-, MOST 107-2634-F-009-006-, and MOST 108-2634-F-009-006-.
	\IEEEcompsocthanksitem The authors are with department of Electrical and Computer Engineering, National Chiao Tung University, 300 Hsinchu, Taiwan.\protect \\Corresponding author's e-mail: lichun@cc.nctu.edu.tw.}
}


\maketitle

\begin{abstract}
In this letter, we study the on-demand UAV-BS placement problem for arbitrarily distributed users. This UAV-BS placement problem is modeled as a knapsack-like problem, which is NP-complete. We propose a density-aware placement algorithm to maximize the number of covered users subject to the constraint of the minimum required data rates per user. Simulations are conducted to evaluate the performance of the proposed algorithm in a real environment with different user densities. Our numerical results indicate that for various user densities our proposed solution can service more users with guaranteed data rates compared to the existing method, while reducing the transmit power by 29\%.
\end{abstract}


%
\IEEEpeerreviewmaketitle

\section{Introduction}
\label{intro}
With the advances of communications techniques, \emph{Unmanned Aerial Vehicle mounted Base Stations} (UAV-BS) becomes a promising solution for providing communications services to flash crowds with instant huge volume of traffic~\cite{7918510}~\cite{7510820}~\cite{8269064}. Nevertheless, UAV-BS are operated in a dynamic and resource-limited communications environment, in which a group of nonuniform and distributed users are needed to be served with a certain amount of low data rates in complicated radio propagation but with only very limited battery capacity. An appropriate placement of UAV-BS in the altitude and the horizontal locations can provide a better communication quality with line-of-sight (LoS) propagation path to ground users. 

Thus, the 3-dimension placement of UAV-BS becomes a hot topic in the literature recently.
These 3D placement research works are addressed from the following aspects: radio channel characteristics~\cite{6863654}, network capacity~\cite{DBLP:journals/corr/KalantariSYY17}~\cite{7762053}, transmission power~\cite{123}, etc. 
First, from the viewpoint of UAV radio channel characteristics, the relation between the altitude of a UAV-BS and its optimal coverage was discussed in~\cite{6863654}, where an air-to-ground (ATG) channel model was proposed to take into account of the probabilities of line-of-sight (LoS) and non-line-of-sight (NLoS). 
Second, from the network capacity aspect, in~\cite{DBLP:journals/corr/KalantariSYY17} a backhaul-aware robust 3D placement of UAV-BS was developed to increase the network capacity. 
In~\cite{7756327}, the optimal density issue of deployed drone small cells was investigated by a 3D Poisson point process (PPP) approach in a spectrum sharing environment with the existing traditional cellular networks.  
In~\cite{7762053}, a polynomial-time UAV-BS deployment approach was proposed to minimize the number of UAV-BSs for various user densities.
Last but not the least, from the power saving aspect, the minimum transmit power issue of a UAV-BS to cover all UEs was addressed in~\cite{123}. 

Rather than addressing the problem of UAV-BS placement from the aspects mentioned in existing literature, we discuss how to deploy a UAV-BS under a unique consideration -- one that recognizes the demands (or traffic requirements) and density of UEs in a serving area.
In this work, we propose a on-demand density-driven UAV 3D placement algorithm to maximize the number of served users.
The addressed 3D placement problem is modeled as a $0$--$1$ knapsack problem, while meeting the data rate requirement of each user and satisfying the constraints of UAV capacity and UAV locations. The detailed problem formulation will be discussed later in Section~\ref{sec:3dplacement}.
In our experiment, we consider a real environment with different UE densities. Our numerical results indicate that the proposed on-demand placement can support the number of users with various user densities, while improving the transmit power efficiency of a UAV-BS by about 29\% .

\section{System Model}
\label{systemmodel}
\begin{figure}[ht]
	\vspace{-13pt}
	\centering
	\includegraphics[width=0.35\textwidth]{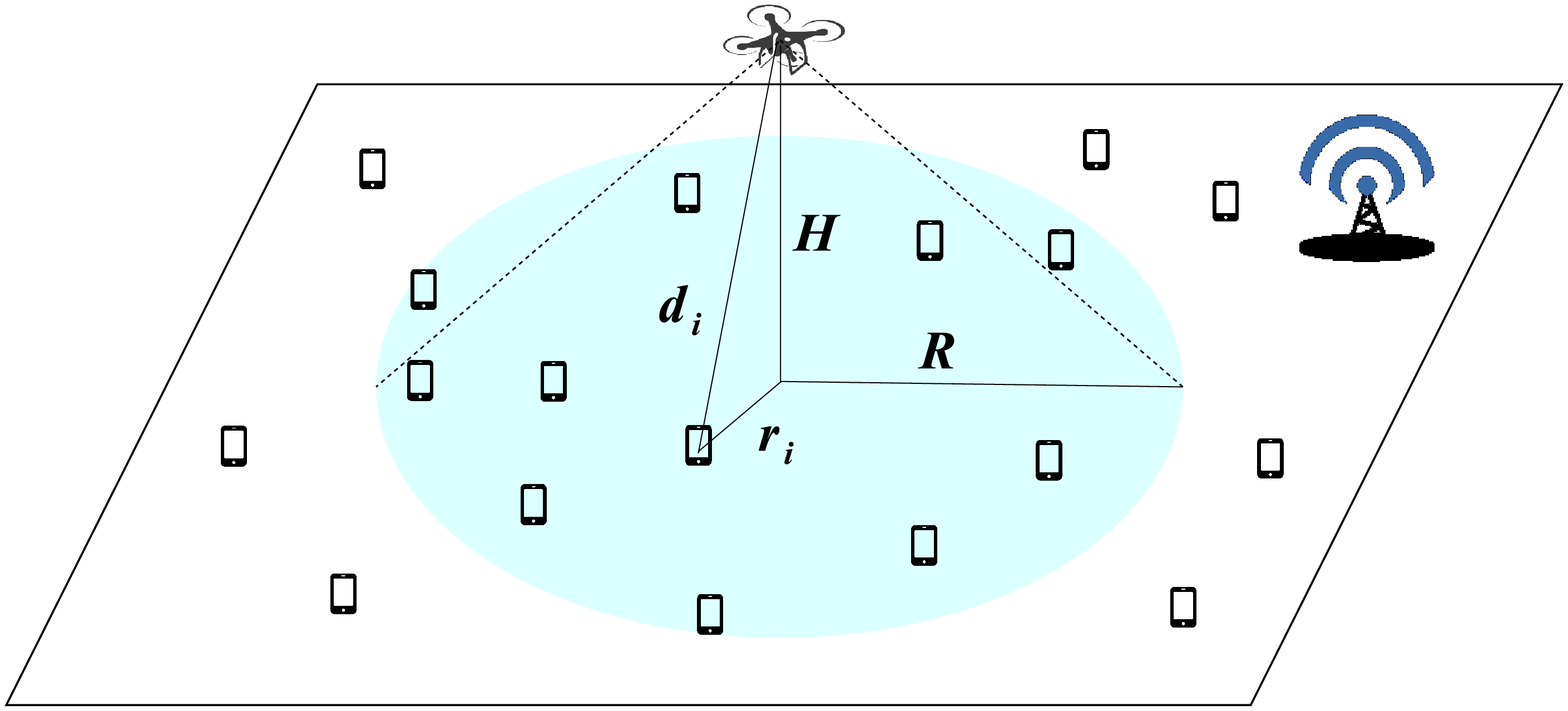}
	\vspace{-5pt}
	\caption{The considered system model.}
	\label{fig:SystemModel}
	\vspace{-5pt}
\end{figure}

Fig.~\ref{fig:SystemModel} illustrates the considered system model in the serving areas, where certain activities, such as concerts, new year activities, marathons, and so on, are held. These events lead to the rapid increase in the number of UEs. Therefore, the ground BS cannot serve so many UEs. Hence, a UAV-BS is dispatched to improve coverage area and data rates for these users. It is assumed that UAV-BS and ground-BS use different spectra, so they do not interfere with each other. We focus on deploying a UAV-BS in a serving area to improve the downlink communication. 


Referring to~\cite{6863654},  we adopt the ATG channel model with the probabilities of LoS and NLoS for a UE as follows:

\begin{align}\label{PLos_PNLoS_to_user}
\text{P}_\text{LoS}(H,r_i)&=\dfrac{1}{1+a\exp(-b(\dfrac{180}{\pi}\tan^{-1}(\dfrac{H}{r_i})-a))},\nonumber\\
\text{P}_\text{NLoS}(H,r_i)&=1-\text{P}_\text{LoS}(H,r_i),
\end{align}
where $H$ is the altitude of UAV-BS in the serving area; $R$ is the coverage of UAV-BS in the serving area, $a$ and $b$ are environment variables; and $r_i=||U-u_i||=\sqrt{(x-x_i)^2+(y-y_i)^2}$ is the horizontal euclidean distance between $u_i$ and $U$. Note that $U=(x,y)$ is the horizontal location of a UAV-BS; $u_i=(x_{i},y_{i})$ is the horizontal location of UE; $E=\{u_1,u_2, \dots ,u_N\}$ is the set of UEs in the serving area; and $|E|=N$ is the total number of UEs. The channel model with LoS and NLoS links can be respectively written as
\begin{align}
\text{L}_\text{LoS}(H,r_i)&=20\log(\dfrac{4\pi f_c d_i}{c})+\eta_{LoS} \nonumber\\
\text{L}_\text{NLoS}(H,r_i)&=20\log(\dfrac{4\pi f_c d_i}{c})+\eta_{NLoS},\label{model_nlos}
\end{align}
where $\eta_{LoS}$ and $\eta_{NLoS}$ are the mean additional losses for LoS and NLoS; $c$ is the speed of light; and $d_i=\sqrt{r_i^2+H^2}$ is the euclidean distance between a UE and UAV-BS. 

According to~\eqref{PLos_PNLoS_to_user} and~\eqref{model_nlos}, and letting $\theta  = {\tan ^{ - 1}}\frac{{{H}}}{{{r_i}}}$,  the average path loss  between $u_i$ and UAV-BS can be calculated as follows:
\begin{align}\label{average_atg_model}
{\text{L}}(H,r_i)&=\text{P}_\text{LoS}(H,r_i)\text{L}_\text{LoS}(H,r_i)+\text{P}_\text{NLoS}(H,r_i)\text{L}_\text{NLoS}(H,r_i) \nonumber\\
&=\dfrac{\eta_{LoS}-\eta_{NLoS}}{1+a\exp(-b(\dfrac{180}{\pi}\theta _i-a))} \nonumber\\
&+20\log({r_i \sec(\theta _i)})+20\log(\dfrac{4\pi f_c}{c})+\eta_{NLoS}.
\end{align}

\section{On-Demand UAV-BS 3D Placement Problem}
\label{sec:3dplacement}
In this work we focus on finding the  horizontal deployment location for UAV since the relation between the altitude and the coverage of a UAV-BS has been discussed in~\cite{6863654}. The problem to be addressed is to find the appropriate UAV-BS location and its radius so as to guarantee the allocated data rate for all the ground users.

$\delta_i\in\{0,1\}$ is an indicator function to represent whether the UE $u_i$ is in the coverage of the UAV-BS. $\delta_i=1$ when $u_i$ is served by UAB-BS ; otherwise, $\delta_i=0$. This necessary condition of $\delta_i=1$ for $u_i$ can be written as 
\begin{equation}\label{indicate_f1}
||U-u_i||\leq R.
\end{equation}
To converge the constraint~\eqref{indicate_f1} for $\delta_i=0$, we rewrite it as~\cite{7918510} 
\begin{equation}\label{indicate_f2}
||U-u_i||\leq R+M(1-\delta_i),
\end{equation}
where $M$ is a large constant to make \eqref{indicate_f2} hold. If $\delta_i=1$, the constraint~\eqref{indicate_f2} will reduce to~\eqref{indicate_f1}. 
After obtaining the coordinate $U$ and the coverage radius $R$ of the UAV-BS, the UAV-BS can allocate the available data rate to the served UEs. This capacity allocation problem is a NP-complete \emph{knapsack-like problem} and can be expressed as 
\begin{align}\label{no_served_ue}
&\hspace{.3em}\max\hspace{.4em}\sum_{i=1}^{N}\delta_i\\
\text{s.t.}&\sum_{i=1}^{N}c_i\delta_i\leq C, \hspace{0.6em}\delta_i\in \{0,1\},\nonumber
\end{align}
where $c_i$ is the allocated data rate of $u_i$ and $C$ is the available data rate of the UAV-BS.

Unlike the existing work to  maximize the number of served UEs, we further take into account of each UE's expected data rate. The UE demand constraints are denoted as $S_{\text{demand}}=\{s_1,s_2,\dots,s_K\}$,  and $K$ is the number of the  levels for guaranteed data rates.

Therefore, the \emph{on-demand UAV-BS placement problem}~\eqref{no_served_ue}  can be expressed as
\begin{align}\label{QoS}
&\hspace{.3em}\max_{U,R}\hspace{.4em}\sum_{i=1}^{N}\delta_i\\
\text{s.t.}&\hspace{.5em}\Phi1:\sum_{i=1}^{N}c_i\delta_i\leq C, \hspace{0.6em}\delta_i\in \{0,1\},\nonumber\\
&\hspace{.5em}\Phi2:c_i\delta_i \geq s_j\delta_i, \hspace{0.6em}\delta_i\in \{0,1\}, j\in\{1,2,\dots,K\},\nonumber\\
&\hspace{.5em}\Phi3:||U-u_i||\leq R+M(1-\delta_i), \nonumber\\ &\hspace{2.6em}\delta_i\in\{0,1\}, \forall u_i\in E,\nonumber \\
&\hspace{.5em}\Phi4:R\leq {R_{max}},\nonumber
\end{align}
where $R_{max}$ is the maximum coverage radius obtained by~\eqref{average_atg_model} with the predefined altitude $H$ of the UAV-BS. 
Our objective is to maximize the number of served UEs, while satisfying the data rate requirement, $s_j$. 
$\Phi1$ is to ensure that the allocated sum rate will not exceed the available data rate, $C$, provided by the UAV-BS. 
$\Phi2$ is to guarantee that the allocated data rate of each UE, $c_i$, meets the data rate requirement, $s_j$, 
where $\delta_i$ makes $\Phi1$ only consider the covered UEs and $j\in\{1,2,\dots,K\}$.
According to $\Phi3$  from~\eqref{indicate_f2}, the system can find some candidate settings including the locations of the UAV-BS, $U$, and the appropriate coverage radius, $R$. $\Phi4$ is the restrictive condition for $\Phi3$ to ensure that the obtained candidate coverage radius, $R$, does not exceed the maximum coverage radius, $R_{max}$, provided by the UAV-BS. 

\vspace{-5pt}
\section{Density-Aware 3D Placement Algorithm for On-Demand UAV-BS Services}
In this section, we present the UAV-BS placement algorithm for provisioning the on-demand instant services to a group of users. 
The proposed density-aware 3D UAV-BS placement algorithm aims to maximize the number of
the arbitrarily distributed users with guaranteed data rates.  
Algorithm~\ref{alg:QoS} shows the proposed placement procedures, which are explained  as follows. 
\begin{enumerate}
	\item The required input information includes $E, s_j, C, R_{max}$, and $m$, where $m$ is the number of iterations for finding candidates. The outcomes of this algorithm are $U=(x,y)$ and $R$. 
	\item Two temporary lists, $D$ and $D'$ are created to store the candidates that may be selected as the output coverage radius. This procedure also results in a temporary array, $r$, to store the horizontal distance between the current $U$ and each $u_i$, where $i = 1, \dots, N$.
	\item  This procedure selects three random UE locations $u_1,u_2,u_3\in E$, and uses the circumcenter to obtain the circumcircle. 
	\item  Let the circumcenter be the UAV-BS location, $U$, and $R$ be the distance from the circumcenter to the selected UEs. 
	\item Runs $m$ iterations to find all the possible placements of UAV-BS from Line~\ref{alg:QoS:line5} to Line~\ref{alg:QoS:line12}.
	\item Lines~\ref{alg:QoS:line6} to~\ref{alg:QoS:line7} determines the distance from the current UAV-BS, $U$, to all UEs, respectively. 
	\item At Line~\ref{alg:QoS:line9}, after calculating $\text{\bf max}(r)$ and $\text{\bf min}(r)$, 
	this algorithm searches the fourth UE $u_4$ such
	that the distance between current $U$ and $u_4$ is the closest to $(\text{\bf max}(r)+\text{\bf min}(r))/2$. 
	\item At Line~\ref{alg:QoS:line10}, the system obtains four possible coverage radii from the combinations of the selected three UEs from $u_1,u_2,u_3$ and $u_4$. Then these candidates are stored in $D'$, where $|D'|=4$.
	\item In the loop at Line~\ref{alg:QoS:line11},  each data $d$ in $D'$ is checked.   
	According to~\eqref{indicate_f2} and~ \eqref{no_served_ue}, 
	$\sum_{i=1}^{N}\delta_i$ is updated by substituting $R$ of each candidate data $d$ in a inner-loop. 
	If $R\le R_{max}$, $c_i\geq s_j$, and $\sum_{i=1}^{N}c_i\delta_i\leq C$, the candidate data $d\leftarrow(x, y, R, c_i, \sum_{i=1}^{N}\delta_i)$ into the list $D$ are saved.	
	\item Finally, after finding the candidate locations and coverage radii, the data entry $d'$ with the maximum $\sum_{i=1}^{N}\delta_i$ in the list $D$ are determined. Then the algorithm outputs $(x, y, R)$ from $d'(x, y, R)$. 
\end{enumerate}

\begin{algorithm2e}[t]
	\small
	\SetAlgoVlined
	\caption{The Proposed On-Demand Density-Aware 3D Placement Algorithm}
	\label{alg:QoS}
	\KwIn{$E, s_j, C, R_{max}, m$}
	\KwOut{$(x, y), R$}
	Create two temporary lists $D'$ and $D$ \label{alg:QoS:line1}\;
	Create a temporary array $r$ and its size is $|E|$ \tcc*[r]{$|E|=N$}	
	Find three UEs, $u_1,u_2,u_3$, derive their circumcenter as the candidate location ($x,y$), and calculate the circumcircle as the candidate coverage radius $R$\label{alg:QoS:line4}\;
	\For{$k=1;k \le m;k++$}{ \label{alg:QoS:line5}
		\For{$i=1;i\le |E|;i++$}{ \label{alg:QoS:line6}     
			$r[i]\leftarrow \sqrt {{{(x - u_i.x)}^2} + {{(y - u_i.y)}^2}}$\label{alg:QoS:line7};\tcc*[f]{$\forall u_i\in E$}	
		}
		Find the fourth UE $u_4$ such that the distance between $U$ and $u_4$ is the nearest one to $(\max(r)+\min(r))/2$ \label{alg:QoS:line9}\;
		Use the combination of three UEs choosing from $u_1,u_2,u_3,u_4$ to obtain four candidate coverages, and store them in $D'$ \label{alg:QoS:line10}\;
		\For{each data entry $d$ in $D'$}{ \label{alg:QoS:line11}
			Update $\sum_{i=1}^{N}c_i\delta_i$ with \eqref{indicate_f2} and \eqref{no_served_ue} by substituting $d.R$\;
			\If{$d.R\le R_{max} \wedge d.c_i \geq s_j \wedge \sum_{i=1}^{N}c_i\delta_i\leq C$}{
				Insert data $d(x, y, R, c_i, \sum_{i=1}^{N}\delta_i)$ into $D$ \label{alg:QoS:line12}\;
			}
		}
	}
	Find $d'$ with the maximum $\sum_{i=1}^{N}\delta_i$ in $D$ and let $(x, y, R)\leftarrow (d'.x, d'.y, d'.R)$ \;
	\Return $(x, y), R$\;
\end{algorithm2e}
\normalsize

Now we analyze the time complexity of  Algorithm~\ref{alg:QoS}.
Since we model the considered on-demand UAV-BS placement problem as a knapsack-like problem with additional constraints, the computational complexity of the optimal placement algorithm increases as the searching space increases. Because the proposed algorithm is designed based on the concept of the genetic algorithm (GA), its complexity can be reduced to polynomial time. The operations from Lines~\ref{alg:QoS:line1} to~\ref{alg:QoS:line4} of Algorithm~\ref{alg:QoS} cost $\mathcal{O}(1)$ time. 
The loop at Line~\ref{alg:QoS:line5} cost $\mathcal{O}(m)$ time and the inner loop at Line~\ref{alg:QoS:line6} needs $\mathcal{O}(N)$ time, where $|E|=N$. 
The operations at Lines~\ref{alg:QoS:line9} and~\ref{alg:QoS:line10} perform the mathematical operations and only need $\mathcal{O}(1)$ time. 
The loop at Line~\ref{alg:QoS:line11} also takes $\mathcal{O}(1)$ time because $|D'|=4$.
In summary, we can know that the procedures of on-demand density-aware 3D placement can be finished in $\mathcal{O}(mN)$ time.


\vspace{-8pt}
\section{Numerical Results and Discussions}
\label{simulation}
In this section, we discuss the system performance in terms of the allocated data rate of each UE and the transmit power of a UAV-BS against user densities. 
We consider the real campus environment of National Chiao Tung University, Taiwan as shown in Fig. 2, where UE are arbitrarily distributed on the campus. 
In this figure, the blue-dashed circle, denoted as $s_{max}$, shows the coverage radius of the UAV-BS with $R_{max}=241.87$ (m) using the existing method~\cite{7918510}.
The red-dashed line circle, $R_4=145.99$ (m), represents the coverage radius of the UAV-BS using the proposed algorithm with the data rate constraint $s_4$. 
The dark-red-dashed line circle, $R_3=83.41$ (m), is the coverage radius using~\eqref{QoS} with the data rate constraint $s_3$.
The orange-dashed circle, $R_2=38.72$ (m), depicts that the coverage radius by the method of~\eqref{QoS} with the data rate constraint $s_2$ 
With the data rate constraint $s_1$, the proposed algorithm obtains deployed coverage radius $R_{1}=24.93$ (m) as shown as the green-dashed circle. 
In this environment, ($a$, $b$, $\eta_{LoS}$, $\eta_{NLoS}$) for the ATG channel model are (4.88, 0.43, 0.1, 21) as adopted for an suburban environment in~\cite{7510820}.
In our simulation, we consider four delivered data rates $s_1$, $s_2$, $s_3$, and $s_4$, where
$s_1 = 4\times10^6$ (bps), $s_2=2\times10^6$ (bps), $s_3=1\times10^6$ (bps), and $s_4=5\times10^5$ (bps) are defined for full HD video streaming, online gaming, web surfing, and VoIP, respectively. 
\begin{figure}[h]
	\centering  
	\vspace{-10pt}
		\includegraphics[width=0.48 \textwidth]{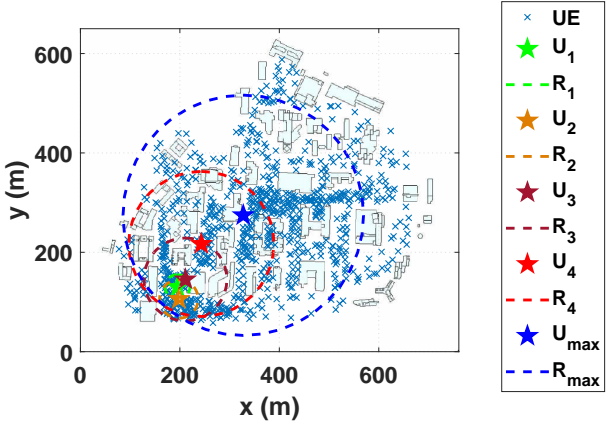}
	\vspace{-5pt}
	\caption{The results of deploying location \& coverage calculated by the proposed algorithm with different data rate constraints, $s_1=4\times10^6$ (bps), $s_2=2\times10^6$ (bps), $s_3=1\times10^6$ (bps), and $s_4=5\times10^5$ (bps), where $s_{max}$ is aimed to maximize number of served UEs and $\lambda(t)=0.0018$ (UEs/m$^2$).}
	\label{fig:QoS_Rmax:coverage}
	\vspace{-5pt}
\end{figure}

The transmit power of a UAV-BS to serve the user at the cell edge is an important performance metric for 3D UAV-BS placement algorithms. According to Shannon capacity theorem, the transmit power $\text{P}_i(r_i)$ for delivering capacity $c_i$ to UE $u_i$ can be calculated by
\begin{equation}\label{channel_capacity}
\text{P}_i(r_i)=10^{{\text{L}}(H,r_i)/10}N_0 b_i (2^{{c_i}/b_i}-1),\nonumber
\end{equation}
where $b_i$ is the allocated bandwidth and path loss $\text{L}(H, r_i)$ is defined in~\ref{systemmodel}.
($N_0$, $f_c$, $C$, $B$) are ($-174$ dBm, $2\times10^9$ Hz, $2\times10^8$ bps, $2\times10^7$ Hz), where $N_0$ is the noise power spectrum density, $f_c$, $C$, and $B$ have been defined in Section~\ref{systemmodel}. Given $c_i$ and $b_i$, 
the expected transmit power of all UEs of UAV-BS~\cite{8269064} can be expressed as
\begin{equation}\label{expected_transmit_power_uav}
\text{P}_t(H,R,\lambda(t))=\lambda(t)\int_0^{R}2\pi r_i\text{P}_i(r_i)dr_i,\nonumber
\end{equation}
where $\lambda(t)$ is the UE density.
Referring to~\cite{6863654} and   \eqref{average_atg_model}, we consider $\text{L}(H, r_i) = 100$ (dB), $R_{max}=241.87$ (m) and $H = 30$ (m) in our simulation platform.

\begin{figure}[h]
	\centering
	\includegraphics[width=0.43\textwidth]{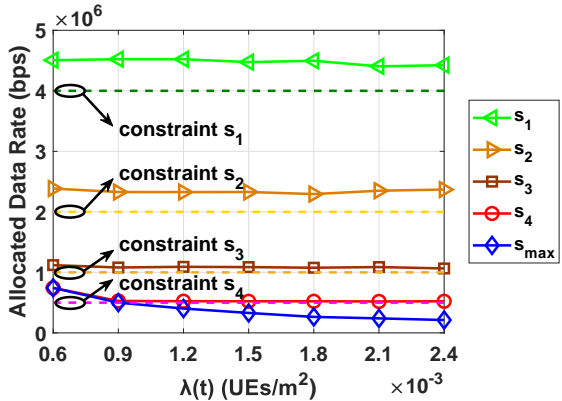}
	\vspace{-5pt}
	\caption{The effect of $\lambda(t)$ on the allocated data rate}
	\label{fig:density:rate}
	\vspace{-5pt}
\end{figure}

Fig.~\ref{fig:density:rate} shows that the proposed approach can always guarantee that the allocated data rate of each UE is higher than the predefined data rate constraint. By contrast, with the strategy~\cite{7918510} of using the case of $s_{max}$, a UAV-BS can cover the largest number of UEs, but such a way may leads resource contention problem and makes the allocated data rate of each UE drop dramatically as the UE density increases. Fig.~\ref{fig:density:rate} indicates that the case of $s_{max}$ only can meets data rate constraint $s_4$ when $\lambda(t)\leq 0.0009$ (UEs/m$^2$). The possible reason is that the existing work~\cite{7918510} is not designed for high density environments. To make it support high density cases, it is suggested that some administration controls be taken, such as association policy, power control, and resource scheduling. 
\vspace{-15pt}

\begin{figure}[h]
	\centering
	\includegraphics[width=0.43\textwidth]{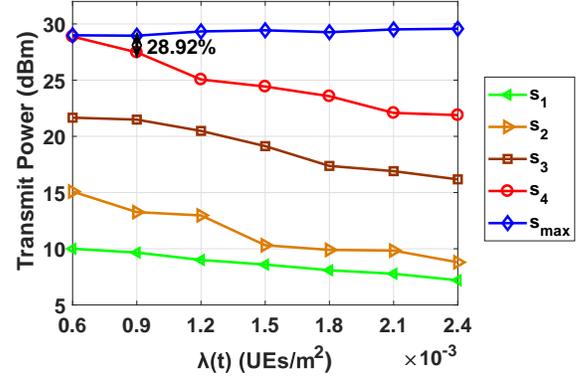}
	\vspace{-5pt}
	\caption{The effect of $\lambda(t)$ on the transmit power}
	\label{fig:density:Lambda_P_R}
	\vspace{-15pt}
\end{figure}

Fig.~\ref{fig:density:Lambda_P_R} indicates that the transmit power of a UAV-BS can be effectively controlled by the proposed approach when $\lambda(t)\geq 0.0009$ (UEs/m$^2$).The reason is that the coverage radius of a UAV-BS is determined with the consideration of UE density by~\eqref{expected_transmit_power_uav}. The transmit power of UAV-BS with the data rate constraint $s_4$ and the case of $s_{max}$ will be almost the same if UE density is too sparse and $\lambda(t)\leq 0.0006$ (UEs/m$^2$). Considering the case of UE density $\lambda(t)=0.0009$ (UEs/m$^2$) with the given UE demand constraint $s_4$, the system will recommend that the coverage radius is 216.85 (m). According the result in Fig.~\ref{fig:density:Lambda_P_R}, the transmit power is improved by more than 29\% in comparison with using the case of $s_{max}$ as UE density increases.
\vspace{-15pt}




\section{Conclusion}
\label{conclusion}


In this letter, we investigate how to deploy a UAV-BS for supporting arbitrarily distributed users 
with various guaranteed data rates. We model the considered on-demand placement problem as a knapsack-like problem. We develop an efficient algorithm to find the place of UAV-BS to 
maximize the served users in polynomial time. In the simulations,  four UE demanded data rates, i.e., full HD video streaming, online gaming, web surfing, and VoIP call or video call , are considered  to verify our proposed method. According to the numerical results, the proposed on-demand placement can guarantee the allocated data rate in all cases. In addition, the transmit power of the UAV-BS can be improved by more than 29\% compared to the existing approach, for which high user density case is  not considered. Many research works are worthwhile being investigated further, such as the interference management for multiple UAV-BSs and the associated deployment strategies. 



\bibliographystyle{ieeetr}
\bibliography{IEEEabrv,WANG_WCL2019-0009}

\end{document}